# Wireless Network Design Under Service Constraints

Martin Kasparick and Gerhard Wunder
Heinrich-Hertz-Lehrstuhl für Informationstheorie und theoretische Informationstechnik
Technische Universität Berlin
Einsteinufer 25, 10587 Berlin, Germany

***Abstract*—In this paper we consider the design of wireless queueing network control policies with special focus on application-dependent service constraints. In particular we consider streaming traffic induced requirements such as avoiding buffer underflows, which significantly complicate the control problem compared to guaranteeing throughput optimality only. Since state-of-the-art approaches for enforcing minimum buffer constraints in broadcast networks are not suitable for application in general networks we argue for a cost function based approach, which combines throughput optimality with flexibility regarding service constraints. New theoretical stability results are presented and various candidate cost functions are investigated concerning their suitability for use in wireless networks with streaming media traffic. Furthermore we show how the cost function based approach can be used to aid wireless network design with respect to important system parameters. The performance is demonstrated using numerical simulations.**

## I. Introduction

Multimedia traffic, in particular streaming content, is increasingly gaining importance as it constitutes an ever larger fraction of the traffic observed in nowadays wireless networks. However, routing streaming content through a wireless network raises hard constraints on the network and corresponding control policies.

Research on queueing control in stochastic networks long focused solely on throughput optimality (i.e. stability), following the celebrated MaxWeight policy of Tassiulas and Ephremides [1]. However it was soon discovered that MaxWeight-type policies can lead to significant delays [2] which may render their application in practice impossible. Significant research activities followed, focusing on delay reduction in backpressure based policies, such as [3][4][5]). A general class of throughput optimal policies with improved delay performance is presented recently in [6], and a general survey on recent policy synthesis techniques can be found for example in [7]. However, delay clearly is not the only performance measure relevant for modern multimedia applications.

In this paper we want to look at wireless network and policy design from a different angle. In particular we adopt a service centric or application centric view point. Our main motivation is that (especially with the ever growing amount of streaming traffic) it is of lesser importance to be e.g. delay optimal as long as the service requirements are fulfilled and–in case of streaming traffic–the stream is not interrupted. In fact, a throughput optimal or even more a delay optimal policy can be even harmful in case of size-limited buffers. In such a case it can be better to absorb traffic peaks at intermediate buffers instead of routing packets as fast as possible to the user. On the other hand, when the application is satisfied with the offered service it is not necessary to put huge efforts in small improvements of particular performance measures. By contrast, it is much more important that a network control policy is able to cope with changing services and thus with changing constraints, however without sacrificing stability.

For control policy synthesis, we assume that the service layer constraints can be expressed as requirements on the buffer states. For example a general delay sensitive application may require certain maximum levels of the buffer states while a multimedia streaming application will additionally need minimum buffer constraints, such that the stream will not be interrupted due to buffer underflows. While the first type of constraint is well investigated, especially in multihop networks the second type bears serious challenges. In a way, this also implies a change of viewpoint: from stabilizing transmit buffers to controlling the user or application buffers[1].

While performance metrics such as throughput, delay (including deadline constrained traffic), and fairness are well investigated in the literature, the problem of minimum buffer constraints was to date mainly considered in the context of stochastic processing networks [8]. Considering queueing networks, previous works consider underflow constraints mainly in broadcast or simpler networks [9][10]. In [9] the problem of guaranteeing minimum buffer constraints in a network of multiple transmitter-receiver pairs with transmitters being able to cooperate is tackled in the following quite simple way. The available resources per user are divided into a fixed and a variable part. Essentially, whenever the user buffer states are low the transmitter increases the variable rate and decreases it when the user buffer states are high. However, this approach is limited to applications in broadcast or point to point links and cannot easily be generalized to the case of arbitrary networks, due to the complex interactions. For example, it is required that buffer levels for each user progress independently of other users ([9]). Clearly this is not the case in a multihop network. Already in the simple case of tandem networks, as depicted in Figure 1, the approach is stretched to its limits when more than two buffers are involved, since all buffers are strongly coupled. Assume we want to steer the level of buffer $m$ towards a certain value, in this case it is not anymore obvious how to control the output for example of the first buffer. This is even further complicated when different costs are assigned to different buffers. Apart from generalization issues to multihop networks, another drawback of these approaches is that they are rather static with respect to service constraints.

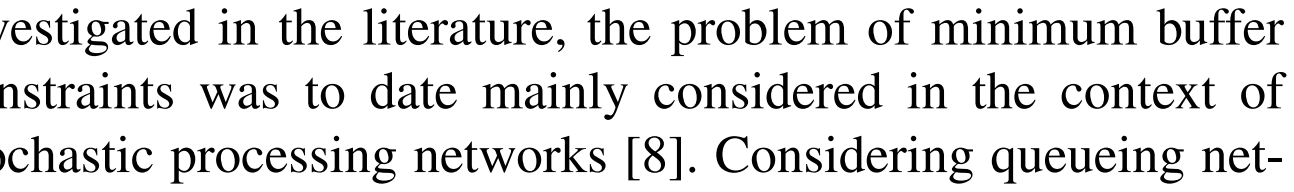

Fig. 1. Tandem Network

[1]Throughout the paper we use the terms 'user buffer' and 'application buffer' synonymously.

The latter is more than a side note since in addition to guaranteeing specific service requirements, a sophisticated control approach should also be flexible with respect to different and possibly changing requirements for different buffers in the network. In [5] we introduced a control framework, called $\mu$-MaxWeight, which–based on an underlying cost function–is highly suitable for application in such cases. Besides, when carefully designed we still maintain throughput optimality. In this paper we provide general sufficient conditions on the stability of a $\mu$-MaxWeight based control policy and subsequently prove two simpler stability results which are based on additional assumptions on the network model.

In summary, our approach guarantees stability of all buffers in the network and, being derived from an underlying cost function, allows incorporation of application requirements. Moreover it can flexibly react to changing user requirements by using different cost functions.

As we will demonstrate, the choice of the underlying cost function is crucial for the performance of the resulting control policy. In particular we show that with a sophisticated cost function choice, buffer sizes of particular queues can be steered towards beneficial operating points. As a main performance measure to compare various cost functions we define a notion of *queue outage* to be minimized. This not only incorporates buffer overflows to capture effects of high packet holding costs or even dropped packets, but also buffer underflows to capture service interruptions.

In addition to above mentioned benefits our framework can also be used to aid a priori design of important network parameters, in order to achieve a given required performance. Eventually–using numerical experiments–we attempt to answer the following two questions:

1) Which cost-function is most suitable to minimize the probability of buffer outage events?
2) How much system resources are needed to achieve a given required outage performance?

As an example for the second issue we consider the problem of wireless bandwidth provisioning.

The rest of the paper is organized as follows. In Section II we introduce our stochastic queueing network model and important stability definitions. In Section III we introduce important preliminaries and in Section IV we introduce our control approach and further present new theoretical results on the stability of resulting policies. Further we propose a pick-and-compare based variant, which significantly reduces complexity. Section V is dedicated to the choice of the underlying cost function and illustrates the implications using a simple tandem queue example. Using numerical simulations, Section VI evaluates the performance of the control approach and exemplarily demonstrates how it can be used for the design of wireless networks. We conclude the paper in Section VII.

*Notation:* We use boldface letters to denote vectors as well as matrices, and common letters with subscript are the elements, such that $A_i$ is the $i$'th element of vector $\boldsymbol{A}$ and $B_{ij}$ is the element in row $i$ and column $j$ of matrix $\boldsymbol{B}$. Moreover $\boldsymbol{A}^T$ refers to the transpose of $\boldsymbol{A}$. $\mathbb{E}\{X\}$ denotes the expected value of random variable $X$. Let $\boldsymbol{I}$ denote the identity matrix of appropriate dimension. Furthermore we denote $\mathbf{1}$ the vector of all ones. $\mathbf{diag}(a_1, a_2, ...)$ refers to a diagonal matrix built from the elements $a_1, a_2, ...$ and $\|\cdot\|_i$ denotes the $l_i$ vector norm and $\|\mathbf{x}\|$ is an arbitrary norm. Furthermore we use $\mathcal{A}^c$ to denote the complement of a set $\mathcal{A}$. The probability of $\mathcal{A}$ is denoted as $\Pr\{\mathcal{A}\}$. The indicator $\mathbb{I}\{\cdot\}$ equals 1 if the argument is true and equals 0 otherwise.

## II. System Model

Similar to [11] we use a simple stochastic network model: the Controlled Random Walk (CRW) model. We consider a queueing network with $m$ queues in total representing $m$ physical buffers with unlimited storage capacity. We arrange the queue backlog in the vector $\mathbf{Q}$, such that $\mathbf{Q} = [Q_1, \ldots, Q_m]^T$ which we refer to as the queue state. Let $\mathcal{M}$ be the set of queue indices. Suppose that the evolution of the queueing system is time slotted with $t \in \mathbb{N}_0$. Then, the CRW model is defined by queueing law:

$$\mathbf{Q}\,(t+1) = \left[\mathbf{Q}\,(t) + \mathbf{B}\,(t+1)\,\mathbf{U}\,(t)\right]^+ + \mathbf{A}\,(t+1) \qquad (1)$$

where $[x]^+ = \max\{0, x\}$. Here, the vector process $\mathbf{A}\,(t) \in \mathbb{N}_0^m$ is the (exogenous) influx to the queueing system with mean $\boldsymbol{\alpha} \in \mathbb{R}_+^m$ (vector of arrival rates in packets per slot); $\mathbf{B}\,(t) \in \mathbb{Z}_0^{m\times l}$ is a matrix process with average $\mathbf{B} \in \mathbb{Z}_0^{m\times l}$, containing both information about network topology (that is, connectivity or routing paths) and service rates. The control $\mathbf{u} = \mathbf{U}\,(t)$ in slot $t$ is an element of the set $\{0,1\}^l$ constrained by $\boldsymbol{C}\boldsymbol{u} \leq \mathbf{1}$ using the binary constituency matrix $\mathbf{C} \in \mathbb{Z}_0^{l_m \times l}$ (with $l_m > 0$ being the number of resource constraints in the network). For the sake of notational simplicity we omit the time index in the following where possible. Throughout the entire paper $\boldsymbol{x} \in \mathbb{N}_0^m$ denotes the current backlog.

In what follows, the queueing system (1) is assumed to be a $\delta_{\mathbf{0}}$-irreducible Markov chain ($\delta_{\mathbf{0}}$ being the point measure at $\boldsymbol{x} = \mathbf{0}$).

### A. Stability

The stability of an $\delta_{\mathbf{0}}$-irreducible Markov chain can be defined in different manners. Throughout this paper we use the following definition of stability.

**Definition 1.** *A Markov chain is called* f-stable*, if there is an unbounded function $f : \mathbb{R}_+^m \to \mathbb{R}_+$, such that for any $0 < B < +\infty$ the set $\mathcal{B} := \{\mathbf{x} : f\,(\mathbf{x}) \leq B\}$ is compact, and furthermore it holds*

$$\limsup_{t\to+\infty} \mathbb{E}\left\{f\,(\mathbf{Q}\,(t))\right\} < +\infty. \qquad (2)$$

In the definition the function $f$ is unbounded in all positive directions so that $f\,(\mathbf{x}) \to \infty$ if $\|\mathbf{x}\| \to \infty$. Choosing directly $f\,(\mathbf{x}) = \|\mathbf{x}\|$, Definition 1 is equivalent to the definition of

*strongly stable* [12] which implies weak stability. Clearly, for any $f(\mathbf{x})$ which grows faster than $\|\mathbf{x}\|$, inequality (2) implies that the Markov chain is strongly stable. We call a vector of arrival rates $\boldsymbol{\alpha} \in \mathbb{R}_+^m$ *stabilizable* when the corresponding queueing system driven by some specific scheduling policy is positive recurrent.

A scheduling policy is now called *throughput optimal* if it keeps the Markov chain positive recurrent for any vector of arrival rates $\boldsymbol{\alpha}$ for which a stabilizing policy exists.

## III. Preliminaries

Let us introduce a cost function

$$c : \mathbb{N}_0^m \to \mathbb{R}_+, \mathbf{x} \hookrightarrow c(\mathbf{x}),$$

assigning any queue state a non-negative number. Typically, the goal is to minimize the average cost over a given finite or infinite time period or some discounted cost criterion. The optimal solution to the resulting problems –which in discrete time can be modeled as a *Markov Decision Problem*– can be found by dynamic programming, which is, however, infeasible for large networks. A simple approach to queueing network control is the *myopic or greedy policy*. Such a policy selects the control decision that minimizes the expected cost only for the next time slot.

In [11], a cost function based policy design framework called $h$-MaxWeight is introduced which is a generalization of the MaxWeight policy. Meyn considers a slightly different definition of the CRW model, which is characterized by queueing law:

$$\mathbf{Q}(t+1) = \mathbf{Q}(t) + \mathbf{B}(t+1)\mathbf{U}(t) + \mathbf{A}(t+1) \tag{3}$$

The control $\mathbf{U}(t) \in \mathbb{N}_0^l$ is an element of the region

$$\mathcal{U}^*(\mathbf{x}) := \mathcal{U}(\mathbf{x}) \cap \{0,1\}^l,$$

with

$$\mathcal{U}(\mathbf{x}) := \left\{\mathbf{u} \in \mathbb{R}_+^l : \mathbf{Cu} \leq \mathbf{1}, [\mathbf{Bu} + \boldsymbol{\alpha}]_i \geq 0 \text{ for } x_i = 0\right\}.$$

In the $h$-MaxWeight based control policy, the control vector is derived according to

$$\underset{\boldsymbol{u} \in \mathcal{U}^*(\mathbf{x})}{\arg\min} < \nabla h(\boldsymbol{x}), \boldsymbol{B}\boldsymbol{u} + \boldsymbol{\alpha} > . \tag{4}$$

Thus, the policy is myopic with respect to the gradient of some perturbation $h$ of the underlying cost function. Meyn develops two main constraints on the function $h$: the first requires the partial derivative of $h$ to vanish when queues become empty:

$$\frac{\partial h}{\partial x_i}(\boldsymbol{x}) = 0 \quad \text{if } x_i = 0 \tag{5}$$

Moreover the dynamic programming inequality has to hold for the function $h$:

$$\min_{\boldsymbol{u} \in \mathcal{U}(\boldsymbol{x})} < \nabla h(\boldsymbol{x}), \boldsymbol{B}\boldsymbol{u} + \boldsymbol{\alpha} > \leq -c(\boldsymbol{x})$$

When $h$ is non-quadratic, the derivative condition (5) is not always fulfilled. Therefore a perturbation technique is used where $h(\boldsymbol{x}) = h_0(\tilde{\boldsymbol{x}})$, hence it is a perturbation of a function $h_0$. Two perturbations are proposed: an exponential perturbation with $\theta \geq 1$ given by

$$\tilde{x}_i := x_i + \theta\left(e^{-\frac{x_i}{\theta}} - 1\right),$$

and a logarithmic perturbation with $\theta > 0$ defined as

$$\tilde{x}_i := x_i \log\left(1 + \frac{x_i}{\theta}\right). \tag{6}$$

While the first approach shows better performance in simulations, the stability of the resulting policy depends on the parameter $\theta$ being sufficiently large (determined by the considered network setting). This is overcome by the second perturbation which is stabilizing for each feasible $\theta$, however it comes with the additional constraint

$$\frac{\partial h_0}{\partial x_i}(\boldsymbol{x}) \geq \epsilon x_i, \quad \forall i \in \mathcal{M},$$

which is a significant limitation on the space of functions that can be chosen as $h_0$.

## IV. $\mu$-MaxWeight

In this section, we give generalized sufficient conditions for throughput optimality of the systems (1) and (3). In what follows, we consider scheduling policies of the form

$$\mathbf{u}^*(\mathbf{x}) = \underset{\mathbf{u} \in \mathbb{R}_+^n : \boldsymbol{C}\boldsymbol{u} \leq \mathbf{1}}{\arg\min} \langle \boldsymbol{\mu}(\mathbf{x}), \mathbf{Bu} + \boldsymbol{\alpha} \rangle, \tag{7}$$

where $\boldsymbol{\mu}(\mathbf{x})$ is a vector valued function $\mathbb{R}_+^m \to \mathbb{R}_+^m$, which is called the *weight vector* for some actual queue state $\mathbf{x}$. Note that $\boldsymbol{\mu}$ is reminiscent of a vector field and can thus be interpreted as a *scheduling field* for which we present a stability characterization. Observe that by construction of the policy we can without loss of generality normalize the weight vector as

$$\bar{\boldsymbol{\mu}}(\mathbf{x}) := \frac{\boldsymbol{\mu}(\mathbf{x})}{\|\boldsymbol{\mu}(\mathbf{x})\|_1} \tag{8}$$

and hence $\|\bar{\boldsymbol{\mu}}(\mathbf{x})\|_1 = 1$. Furthermore, we assume that the resulting policy is non-idling, i.e. $\|\boldsymbol{\mu}(\mathbf{x})\|_1 = 0$ if and only if $\mathbf{x} = \mathbf{0}$.

**Theorem 1.** *Consider the queueing system (1) driven by the control policy (7) with some scheduling field $\boldsymbol{\mu}$. The policy is throughput optimal if the corresponding normalized scheduling field given in Eqn. (8) fulfills the following conditions:*

1) *Given any $0 < \epsilon_1 < 1$ and $C_1 > 0$, there is some $B_1 > 0$ so that for any $\Delta\mathbf{x} \in \mathbb{R}^m$ with $\|\Delta\mathbf{x}\| < C_1$, we have $|\bar{\mu}_i(\mathbf{x} + \Delta\mathbf{x}) - \bar{\mu}_i(\mathbf{x})| \leq \epsilon_1$ for any $\mathbf{x} \in \mathbb{R}_+^m$ with $\|\mathbf{x}\| > B_1$, $\forall i \in \mathcal{M}$.*
2) *Given any $0 < \epsilon_2 < 1$ and $C_2 > 0$, there is some $B_2 > 0$ so that for any $\mathbf{x} \in \mathbb{R}_+^m$ with $\|\mathbf{x}\| > B_2$ and $x_i < C_2$, we have $\bar{\mu}_i(\mathbf{x}) \leq \epsilon_2$, for any $i \in \mathcal{M}$.*

*Moreover, for any stabilizable arrival process the queueing system is f-stable under the given policy where $f$ is an unbounded function as defined in Definition 1. The exact formulation of $f$ depends on the field $\bar{\boldsymbol{\mu}}(\mathbf{x})$.*

*Proof:* The proof is based on [5] and is provided in full detail in [13]. ■

With further assumptions on the underlying network these conditions can be significantly simplified. The following additional requirement on the network topology (also assumed in Theorem 1.1 of [11]) is needed for subsequent corollaries.

- $B_{ij}(t) \geq -1$ for each $i$,$j$ and $t$, and for each $j \in \{1, \ldots, l_u\}$ there exists a unique value $i_j \in \{1, \ldots, l\}$ satisfying
$$B_{ij}(t) \geq 0 \quad \text{a.s. } \forall i \neq i_j. \tag{9}$$

**Corollary 1.** *Consider the queueing system (3) driven by the control policy (4) with some cost function $h$. Suppose the corresponding scheduling field $\boldsymbol{\mu}(\boldsymbol{x}) := \nabla h(\boldsymbol{x})$ is continuously differentiable and Condition (9) on network topology $\{\boldsymbol{B}(\cdot)\}$ holds. Then, the following conditions are sufficient for throughput optimality:*

1) *For any $\epsilon > 0$ there is some $C_1^* > 0$ so that for all $\|\boldsymbol{x}\| \geq C_1^*$:*
$$\|\nabla \log(\mu_i(\boldsymbol{x}))\| \leq \epsilon, \quad \forall i \in \mathcal{M}$$
2) *If $x_i = 0$ then $\mu_i(\boldsymbol{x}) = 0$, $\forall i \in \mathcal{M}$.*

*Proof:* By Condition 2) of Corollary 1 we can assume that the random walk evolves on $\mathbb{R}_+^m$. Hence, we can skip Condition 2) of Theorem 1 since this condition (as its counterpart in Corollary 1) ensures positivity of the random walk. We need to show that from

$$\|\nabla \log \mu_i(\boldsymbol{x})\| \leq \epsilon, \quad \forall i \in \mathcal{M}, \ \|\boldsymbol{x}\| > C_6(\epsilon), \tag{10}$$

(where $C_6(\epsilon)$ is sufficiently large) it follows:

$$\left| \frac{\mu_i(\boldsymbol{x} + \Delta\boldsymbol{x})}{\sum_{j \in \mathcal{M}} \mu_j(\boldsymbol{x} + \Delta\boldsymbol{x})} - \frac{\mu_i(\boldsymbol{x})}{\sum_{j \in \mathcal{M}} \mu_j(\boldsymbol{x})} \right| \leq \epsilon \tag{11}$$

For orientation, let us assume more restrictive conditions first: take $\mu_i$, $\forall i \in \mathcal{M}$ Lipschitz continuous and let $\sum_{j \in \mathcal{M}} \mu_j(\boldsymbol{x}) \to \infty$ if $\|\boldsymbol{x}\| \to \infty$. Note, that these conditions already encompasses Meyn's perturbation (6) together with e.g. a linear cost function.

It is easy to prove the corollary with these assumptions: by the mean value theorem we have

$$\mu_i(\boldsymbol{x} + \Delta\boldsymbol{x}) = \mu_i(\overline{\boldsymbol{x}}) + \nabla_{\boldsymbol{x}}^T \mu_i(\widetilde{\boldsymbol{x}}) \Delta\overline{\boldsymbol{x}}$$

where $\overline{\boldsymbol{x}}$ is an (arbitrary) point on line connecting $\boldsymbol{x}$ and $\boldsymbol{x}+\Delta\boldsymbol{x}$ whereas $\widetilde{\boldsymbol{x}}$ is a point connecting $\overline{\boldsymbol{x}}+\Delta\overline{\boldsymbol{x}}$. Since the field is Lipschitz we have $\nabla_{\boldsymbol{x}}^T \mu_i(\overline{\boldsymbol{x}}) \leq C_7$ uniformly. Furthermore, since the policy is non-idling $\sum_{j \in \mathcal{M}} \mu_j(\boldsymbol{x} + \Delta\boldsymbol{x}) \geq C_8$ where the normalization constant $C_8$ can be chosen as large as possible without altering the policy (by the construction of the policy). Moreover, since $\sum_{j \in \mathcal{M}} \mu_j(\boldsymbol{x}) \to \infty, \|\boldsymbol{x}\| \to \infty$, condition (11) is equivalent to

$$|\mu_i(\boldsymbol{x} + \Delta\boldsymbol{x}) - \mu_i(\boldsymbol{x})| \leq \epsilon \sum_{j \in \mathcal{M}} \mu_j(\overline{\boldsymbol{x}})$$

and, again, by the mean value theorem:

$$\left|\nabla^T \mu_i(\overline{\boldsymbol{x}}) \Delta\boldsymbol{x}\right| \leq \epsilon \sum_{j \in \mathcal{M}} \mu_j(\overline{\boldsymbol{x}})$$

Here, we tacitly assumed that we have selected $\overline{\boldsymbol{x}}$ accordingly. Since $\Delta\boldsymbol{x}$ is fixed and by the positivity of $\mu_i$ it is sufficient that

$$\|\nabla \mu_i(\boldsymbol{x})\| \leq \frac{\epsilon}{\|\Delta\boldsymbol{x}\|} \mu_i(\boldsymbol{x})$$

which is equivalent to condition (10) with some $\|\boldsymbol{x}\| > C_6(\epsilon')$ ($\epsilon'$ slightly smaller).

Let us now prove the general case. Condition (10) can be written as

$$\frac{1}{\sum_{j \in \mathcal{M}} \mu_j(\boldsymbol{x})} \nabla^T \mu_i(\boldsymbol{x}) \Delta\boldsymbol{x} = \epsilon_n,$$

for some $\boldsymbol{x}$ with $\|\boldsymbol{x}\| > C(\epsilon_n)$ where $\epsilon_n$ is a zero sequence and $C(\epsilon_n)$ is strictly increasing for any fixed $\Delta\boldsymbol{x} \in \mathbb{R}^m$. Now, again, by the mean value theorem

$$\left| \frac{\mu_i(\boldsymbol{x} + \Delta\boldsymbol{x})}{\sum_{j \in \mathcal{M}} \mu_j(\bar{\boldsymbol{x}}) + \nabla^T \mu_j(\tilde{\boldsymbol{x}}) \Delta\bar{\boldsymbol{x}}} - \frac{\mu_i(\boldsymbol{x})}{\sum_{j \in \mathcal{M}} \mu_j(\bar{\boldsymbol{x}}) - \nabla^T \mu_j(\underline{\tilde{\boldsymbol{x}}}) \Delta\underline{\boldsymbol{x}}} \right| \leq \epsilon, \tag{12}$$

where we set $\bar{\boldsymbol{x}}$ as before and let $\boldsymbol{x} + \Delta\underline{\boldsymbol{x}} = \bar{\boldsymbol{x}}$ and $\bar{\boldsymbol{x}} + \Delta\bar{\boldsymbol{x}} = \boldsymbol{x} + \Delta\boldsymbol{x}$. $\underline{\tilde{\boldsymbol{x}}}, \tilde{\boldsymbol{x}}$ are points on the line connecting $\boldsymbol{x}$ and $\bar{\boldsymbol{x}}$ respectively $\bar{\boldsymbol{x}}$ and $\boldsymbol{x} + \Delta\boldsymbol{x}$. Note that $\mu_j(\bar{\boldsymbol{x}})$ is zero if and only if $\mu_i(\boldsymbol{x} + \Delta\boldsymbol{x})$ and $\mu_i(\boldsymbol{x})$ are both zero since otherwise by condition (10) the gradient would be zero as well. Since in this case the condition is trivially satisfied so that we exclude it.

Hence from (12) it follows

$$\left| \mu_i(\boldsymbol{x} + \Delta\boldsymbol{x}) - \mu_i(\boldsymbol{x}) \frac{\sum_{j \in \mathcal{M}} \mu_j(\bar{\boldsymbol{x}})(1 + \overbrace{\frac{\nabla^T \mu_j(\tilde{\boldsymbol{x}}) \Delta\bar{\boldsymbol{x}}}{\mu_j(\bar{\boldsymbol{x}})}}^{(A)})}{\sum_{j \in \mathcal{M}} \mu_j(\bar{\boldsymbol{x}})(1 - \underbrace{\frac{\nabla^T \mu_j(\underline{\tilde{\boldsymbol{x}}}) \Delta\underline{\boldsymbol{x}}}{\mu_j(\bar{\boldsymbol{x}})}}_{(B)})} \right|$$
$$\leq \epsilon \cdot \sum_{j \in \mathcal{M}} \mu_j(\bar{\boldsymbol{x}}) \left(1 + \frac{\nabla^T \mu_j(\tilde{\boldsymbol{x}}) \Delta\bar{\boldsymbol{x}}}{\mu_j(\bar{\boldsymbol{x}})}\right).$$

We can prove that, because of condition (10), (A) and (B) are zero sequences: suppose $\nabla^T \mu_j(\tilde{\boldsymbol{x}})$ is non-zero (then we can stop anyway) then by the repeated application of the mean value theorem, the denominator of, say, (A) can be written as:

$$\mu_j(\bar{\boldsymbol{x}}) = \mu_j(\tilde{\boldsymbol{x}}) + \nabla \mu_j(\boldsymbol{x}_2) \Delta\boldsymbol{x}_2$$

This process generates sequences in $\mathbb{R}_+^m$ with $\tilde{\boldsymbol{x}} = \boldsymbol{x}_1, \boldsymbol{x}_2, ...$ and $\Delta\bar{\boldsymbol{x}} = \Delta\bar{\boldsymbol{x}}_1 \subset \Delta\bar{\boldsymbol{x}}_2, ...$ which are bounded and hence we can pick subsequences converging to some set of limit points $\boldsymbol{x}_\infty^{(k)}, k = 1, 2, ....$ Note that we can restrict the number of limit points to at most two since by definition every limit point is visited arbitrarily often and infinitely close and by construction of the sequence there is no possibility of more than two limit points which neither contain the other in between them. Take these two limit points with corresponding subsequence $\boldsymbol{x}_n^{(k)}, k = 1, 2$: by continuous differentiability we have $\mu_j(\boldsymbol{x}_n^{(k)}) \to \mu_j(\boldsymbol{x}_\infty^{(k)})$ and $\nabla \mu_j(\boldsymbol{x}_n^{(k)}) \to \nabla \mu_j(\boldsymbol{x}_\infty^{(k)}), k = 1, 2$. It must also hold in the limit:

$$\mu_j(\boldsymbol{x}_\infty^{(1)}) + \nabla^T \mu_j(\boldsymbol{x}_\infty^{(2)})(\boldsymbol{x}_\infty^{(1)} - \boldsymbol{x}_\infty^{(2)}) = \mu_j(\boldsymbol{x}_\infty^{(2)})$$

(and vice versa). Since then

$$\frac{\nabla^T \mu_j(\boldsymbol{x}_\infty^{(2)})(\boldsymbol{x}_\infty^{(1)} - \boldsymbol{x}_\infty^{(2)})}{\mu_j(\boldsymbol{x}_\infty^{(2)})\mu_j(\boldsymbol{x}_\infty^{(2)})} \leq \epsilon,$$

(and vice versa) where $\epsilon > 0$ is arbitrarily small by condition (10) we conclude that $\mu_j(\boldsymbol{x}_\infty^{(1)}) = \mu_j(\boldsymbol{x}_\infty^{(2)})$ (but not necessarily $\boldsymbol{x}_\infty^{(1)} = \boldsymbol{x}_\infty^{(2)}$).

Now, we can proceed the process sufficiently often as

$$\begin{aligned} &\frac{\nabla^T \mu_j(\boldsymbol{x}_1)\Delta\boldsymbol{x}_1}{\mu_j(\bar{\boldsymbol{x}})} \\ &\leq \frac{\nabla^T \mu_j(\boldsymbol{x}_1)\Delta\boldsymbol{x}_1}{\mu_j(\boldsymbol{x}_1)\left(1 + \frac{\nabla^T \mu_j(\boldsymbol{x}_2)\Delta\boldsymbol{x}_2}{\mu_j(\boldsymbol{x}_1)}\right)} \\ &\leq \dots \end{aligned}$$

such that in the final step

$$\begin{aligned} \frac{\nabla^T \mu_j(\boldsymbol{x}_{n+1})\Delta\boldsymbol{x}_{n+1}}{\mu_j(\boldsymbol{x}_n)} &= \frac{(\nabla^T \mu_j(\boldsymbol{x}_\infty^{(k)}) + \epsilon_n^k)\Delta\boldsymbol{x}_{n+1}}{\mu_j(\boldsymbol{x}_\infty^{(l)}) + \epsilon_n^l} \\ &= \frac{(\nabla^T \mu_j(\boldsymbol{x}_\infty^{(k)}) + \epsilon_n^k)\Delta\boldsymbol{x}_{n+1}}{\mu_j(\boldsymbol{x}_\infty^{(k)}) + \epsilon_n^l} \\ &\leq \epsilon,\ k,l = 1,2, \end{aligned}$$

by condition (10). Hence, we have

$$\begin{aligned} \frac{\sum_{j\in\mathcal{M}} \mu_j(\bar{\boldsymbol{x}})\left(1 + \frac{\nabla^T \mu_j(\bar{\boldsymbol{x}})\Delta\bar{\boldsymbol{x}}}{\mu_j(\bar{\boldsymbol{x}})}\right)}{\sum_{j\in\mathcal{M}} \mu_j(\bar{\boldsymbol{x}})\left(1 + \frac{\nabla^T \mu_j(\bar{\boldsymbol{x}})\Delta\boldsymbol{x}}{\mu_j(\bar{\boldsymbol{x}})}\right)} &= \frac{(1+\epsilon_n')}{(1+\epsilon_n'')} \\ &= 1 + \epsilon_n''', \end{aligned}$$

$\epsilon_n', \epsilon_n''$ zero sequences, and further

$$\begin{aligned} &|\mu_i(\boldsymbol{x} + \Delta\boldsymbol{x}) - \mu_i(\boldsymbol{x})(1 + \epsilon_n''')| \\ &\leq |\mu_i(\boldsymbol{x} + \Delta\boldsymbol{x}) - \mu_i(\boldsymbol{x})| + \mu_i(\boldsymbol{x})\epsilon_n''' \\ &\leq \epsilon \sum_{j\in\mathcal{M}} \mu_j(\tilde{\boldsymbol{x}})(1 + \epsilon_n'), \end{aligned}$$

which is equivalent to:

$$|\mu_i(\boldsymbol{x} + \Delta\boldsymbol{x}) - \mu_i(\boldsymbol{x})| \leq \epsilon \sum_{j\in\mathcal{M}} \mu_j(\bar{\boldsymbol{x}})(1 + \epsilon_n') - \epsilon_n''\mu_i(\boldsymbol{x}).$$

Since $\bar{\boldsymbol{x}}$ is arbitrary and can be suitably choose, condition (10) with some $\|\boldsymbol{x}\| > C_6\left(\epsilon''''\right)$ is sufficient for the latter to hold. ■

**Corollary 2.** *Suppose, everything is as in Corollary 1. Let the scheduling field be defined as $\boldsymbol{\mu}(\boldsymbol{x}) := \nabla h_0(\tilde{\boldsymbol{x}})$ for some given simple perturbation $\tilde{\boldsymbol{x}}$. Then, for some $\epsilon > 0$,*

$$\frac{\partial \tilde{x_i}}{\partial x_i} \text{ is Lipschitz, and } \frac{\partial \tilde{x_i}}{\partial x_i} \to \infty, x_i \to \infty,$$

$$\frac{\partial h_0}{\partial x_i} \text{ is Lipschitz, and } \frac{\partial h_0}{\partial \tilde{x_i}}(\tilde{\boldsymbol{x}}) \geq \left(\frac{\partial \tilde{x_i}}{\partial x_i}\right)^{1+\epsilon}, x_i \to \infty,$$

*is sufficient for stability.*

*Proof:* We can write

$$\mu_i(\boldsymbol{x}) = \frac{\partial h}{\partial x_i}(\boldsymbol{x}) = l(x_i)\frac{\partial h_0}{\partial \tilde{x_i}}(\tilde{\boldsymbol{x}})$$

where we defined $l := \frac{\partial \tilde{x_i}}{\partial x_i}$. Note, that here $\tilde{x_i}$ only depends on $x_i$. The gradient of the weight $\mu_i(\boldsymbol{x})$ is given by:

$$\frac{\partial \mu_i}{\partial x_j}(\boldsymbol{x}) = \begin{cases} \frac{\partial l}{\partial x_i}(x_i)\frac{\partial h_0}{\partial \tilde{x_i}}(\tilde{\boldsymbol{x}}) + l(x_i)\frac{\partial}{\partial x_i}\frac{\partial h_0}{\partial \tilde{x_i}}(\tilde{\boldsymbol{x}}) & i = j \\ \frac{\partial}{\partial x_j}\frac{\partial h_0}{\partial \tilde{x_i}}(\tilde{\boldsymbol{x}}) \cdot l(x_i) & i \neq j \end{cases}$$

Define $\boldsymbol{x}^\Delta := \boldsymbol{x} + \Delta\boldsymbol{x}$ and $\tilde{\boldsymbol{x}}^\Delta := \tilde{\boldsymbol{x}}(\boldsymbol{x}^\Delta)$. From the proof of Corollary 1 it is clear that we only have to show that

$$\frac{\left|\nabla^T \mu_i(\boldsymbol{x})\Delta\boldsymbol{x}\right|}{\|\boldsymbol{\mu}(\boldsymbol{x}^\Delta)\|} \leq \epsilon,$$

for some $\epsilon > 0$ arbitrarily small. This can be rewritten as:

$$\begin{aligned} &\frac{\frac{\partial l}{\partial x_i}(x_i)\frac{\partial h_0}{\partial \tilde{x_i}}(\tilde{\boldsymbol{x}})\Delta x_i + l(x_i)\frac{\partial}{\partial x_i}\frac{\partial h_0}{\partial \tilde{x_i}}(\tilde{\boldsymbol{x}})\Delta x_i}{\sum_{j\in\mathcal{M}} l(x_j^\Delta)\frac{\partial h_0}{\partial \tilde{x_j}}(\tilde{\boldsymbol{x}}^\Delta)} \\ &\qquad + \frac{l(x_i)\sum_{j\in\mathcal{M}, j\neq i}\frac{\partial}{\partial x_j}\frac{\partial h_0}{\partial \tilde{x_i}}(\tilde{\boldsymbol{x}})\Delta x_j}{\sum_{j\in\mathcal{M}} l(x_j^\Delta)\frac{\partial h_0}{\partial \tilde{x_j}}(\tilde{\boldsymbol{x}}^\Delta)} \leq \epsilon \end{aligned}$$

Since $\frac{\partial h_0}{\partial \tilde{x_i}}, l$ are Lipschitz, thus $\frac{\partial}{\partial x_j}\frac{\partial h_0}{\partial \tilde{x_i}}, \frac{\partial l}{\partial x_i}$ are uniformly bounded, and $l(x_i), \frac{\partial h_0}{\partial \tilde{x_i}}(\tilde{\boldsymbol{x}}) \geq l^{1+\epsilon}(x_i) \to \infty$ when $x_i \to \infty$, the effect of $\Delta\boldsymbol{x}$ vanishes in the denominator. The condition $\frac{\partial h_0}{\partial \tilde{x_i}}(\tilde{\boldsymbol{x}}) \geq l^{1+\epsilon}(x_i)$ is required since we have expressions of the form

$$\frac{l(x_i)l(x_j)}{l(x_i)\frac{\partial h_0}{\partial \tilde{x_i}}(\tilde{\boldsymbol{x}}) + l(x_j)\frac{\partial h_0}{\partial \tilde{x_j}}(\tilde{\boldsymbol{x}})}$$

which then become arbitrarily small. ■

### *A. Handling Complexity: A Pick and Compare Approach*

Centralized throughput optimal scheduling policies, such as MaxWeight, usually suffer from a high computational complexity. This naturally also applies to a $\mu$-MaxWeight based policy. More precisely, a large computational burden arises since the selection of the best control vector in (7) has to be carried out in every time slot, since the number of candidate control vectors grows exponentially with the size of the vector. To tackle the complexity issue there are several known approaches, such as randomized pick-and-compare based methods [14] which reduce complexity at the expense of higher delay or Greedy/Maximal scheduling [15] which has good delay performance but achieves only a fraction of the throughput region.

To circumvent the complexity problem we propose a randomized version based on the first approach. As noted for example in [14] and [16], throughput optimality can be preserved by using a linear-complexity randomized Pick-and-Compare method. Tailored to $\mu$-MaxWeight, the approach can be summarized as follows: At $t = 0$, use $\boldsymbol{u}(0) = \hat{\boldsymbol{u}} \in \mathcal{U}^*$ chosen randomly. Afterwards in each timeslot $t > 0$ first pick a control $\hat{\boldsymbol{u}} \in \mathcal{U}^*$ randomly, and second, choose the control $\boldsymbol{u}(t)$ of this particular timeslot to be either the vector $\hat{\boldsymbol{u}}$ if

$$\langle \boldsymbol{\mu}(\boldsymbol{x}), \boldsymbol{B}\hat{\boldsymbol{u}} + \boldsymbol{\alpha} \rangle < \langle \boldsymbol{\mu}(\boldsymbol{x}), \boldsymbol{B}\boldsymbol{u}(\boldsymbol{t}-\boldsymbol{1}) + \boldsymbol{\alpha} \rangle,$$

or $\boldsymbol{u}(t-1)$ otherwise. Above algorithm preserves throughput optimality as long as

$$\mathbb{P}\left(\hat{\boldsymbol{u}} = \boldsymbol{u}^*\right) \geq \delta,$$

for $\delta > 0$ [14][16] (which is trivially satisfied). The reduced complexity, however, comes at the expense of a higher convergence time. Yet, a tradeoff can be achieved by repeatedly applying the pick and compare steps in every particular timeslot. In the simulations presented in Section VI, we apply the procedure $n_s$ times per timeslot.

Note that, although not focus of this paper, the randomized algorithm was shown to be amenable as a basis for implementing decentralized throughout optimal control policies [17][18].

## V. Cost-Function Choice

A vital design choice in the proposed control approach is the underlying cost-function to be minimized. Different applications induce different constraints on network control and thus need different cost functions. As mentioned before, applications may require both minimum and maximum buffer constraints. Assume we want to find a cost function that is best suited to steer the buffer levels of application buffers towards a target buffer state $\tilde{Q}$.

In previous work such as [5][19] a *linear cost function* was used, given by

$$c(\boldsymbol{Q}) = \sum_i c_i Q_i, \tag{13}$$

since the aim was to minimize total buffer occupancy, corresponding to end-to-end delay. However, this cost function is unsuitable to avoid buffer underflows since it does not penalize buffer states below the target level. A simple and straight forward cost function choice that penalizes deviations from target buffer state $\tilde{Q}$ in both directions is the *shifted quadratic cost function*, given by

$$c(\boldsymbol{Q}) = \sum_i c_i (Q_i - \tilde{Q})^2. \tag{14}$$

However (14) naively treats all buffers in the network equally although most likely only application buffers have minimum state constraints.

In fact, a better performance can be observed by combining (13)-(14), such that only user buffers have quadratic cost terms, while all other buffers induce linear costs. We call the resulting function *composite cost function*, given by

$$c(\boldsymbol{Q}) = \sum_{i \in \mathcal{I}_u} c_i (Q_i - \tilde{Q})^2 + \sum_{j \notin \mathcal{I}_u} c_j Q_j, \tag{15}$$

where $\tilde{Q}$ denotes the desired target buffer level of the application buffers[2] and $\mathcal{I}_u$ denotes the set of all user buffer indices.

Of course, the shifted quadratic terms in (15) are only one of many possibilities to steer the buffers towards a desired working point. Intuitively, any cost function that produces low costs around the target level and increasing costs for underflows and overflows should lead to the anticipated behaviour. Another approach is to explicitly design a cost function with desired properties. Motivated by the required behaviour we additionally consider a cost function which is inspired by the transfer function of a Butterworth bandstop filter, which we subsequently call *bandstop cost function*.

The resulting cost function, is given by

$$c(\boldsymbol{Q}) = \sum_{i \in \mathcal{I}_u} c_{\max} \left( 1 - \frac{1}{1 + \left( \frac{Q_i - \tilde{Q}}{w} \right)^k} \right) + \sum_{j \notin \mathcal{I}_u} c_j Q_j. \tag{16}$$

Thereby the parameter $k$ is the analogue to the filter's order, $w$

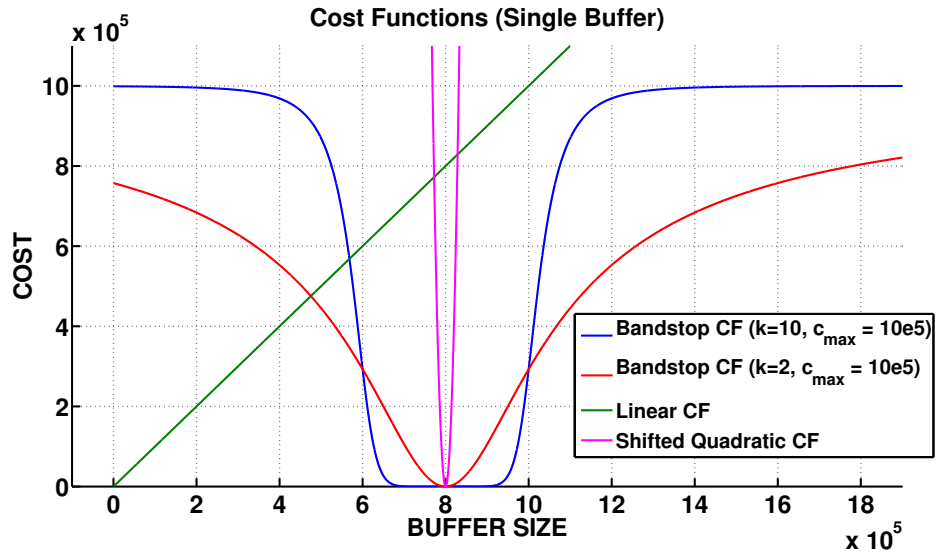

Fig. 2. Cost function trajectories for a single buffer, $\tilde{Q} = 8 \cdot 10^5$

determines the width of the interval of buffer state around the target buffer level that produce very low costs, and $c_{\max}$ scales the function to the desired maximum cost value that occurs when the deviation from the target buffer state is significantly large. For a single buffer system the considered cost functions are illustrated in Figure 2.

Having chosen an appropriate cost function the question remains, how to construct a corresponding weight function $\boldsymbol{\mu}$. In order to guarantee stability, it is sufficient to show that the weight function fulfills the stability conditions of Theorem 1. For this purpose we employ a perturbation technique (cf. [11]). In [5] we pointed out that a simple way to construct a weight function is $\boldsymbol{\mu}(\boldsymbol{x}) = \mathbf{P}_\theta(\mathbf{x}) \nabla h_0(\boldsymbol{x})$, where a *perturbation matrix* $\mathbf{P}_\theta(\mathbf{x}) := \mathbf{diag}\left(1 - \exp\left(-\frac{x_i}{\theta(1+\sum_{j \neq i} x_j)}\right)\right)$ is used. It is based on the following perturbation of variables:

$$\tilde{x}_i := x_i + \exp\left(-\frac{x_i}{\theta(1 + \sum_{j \neq i} x_j)}\right). \tag{17}$$

It can be easily verified that the conditions given in Section IV hold for suitable $h_0$, i.e. it is throughput optimal for any $\theta > 0$ ([5][13]). Here we directly use our cost function as $h_0$. For sufficiently large buffers the stability conditions are clearly fulfilled (while the behaviour when buffers are close to zero is controlled by perturbation (17)).

In general for throughput optimality only the behaviour matters when buffers grow very large [6]. Therefore we are not necessarily confined to everywhere differentiable cost functions. It is also possible to use a for example piecewise linear function to construct the weight function and still maintain throughput optimality, by using an arbitrary subgradient in

[2] For simplicity we assume all application buffers have the same target buffer level.

(7) at non-differentiable points. Although the control that is chosen at these point may not necessarily lead to a negative drift, concerning the stability conditions given in Section IV it is sufficient to assume that the cost function is differentiable when $\|\boldsymbol{x}\|$ is larger than some arbitrary $B$.

### A. Example: Controlling a network of queues in tandem

Let us clarify the influence of the cost-function choice by considering a very simple network known as tandem queue (cf. Figure 1), comprising a number of $m$ buffers in series. To model for example a streaming service, we assume traffic arrives at the first buffer with mean rate $\alpha$ and some application removes traffic from the $m$'th buffer at a constant rate $R_a$. The output of buffers 1 through $m-1$ can be regulated by the control policy. While 'ordinary' tandem queue networks (without considering a specific application) are investigated thoroughly in the literature (see e.g. [7]), we have two additional issues here. First, we have no explicit control over the rate at which data is extracted from application buffer $Q_m$. Second, in addition to stability (i.e. boundedness of buffers from above) we also have minimum buffer state requirements.

Consider now the most simple network of $m = 2$. The second queue is thereby considered as the application queue, thus the only control dimension remaining is whether to send traffic from queue 1 to queue 2 or not. How the buffer state can be steered towards the target buffer level using an appropriate cost function is shown in Figure 3. It depicts queueing trajectories[3] of this system for 20000 time slots using two different underlying cost functions. The dashed lines represent policies based on the simple linear cost function, given in (13). This cost function obviously does not stop the second buffer from growing, since both buffers are weighted equally. By contrast, the solid lines are obtained using the composite cost function (15), which produces a quadratic cost at the second buffer when the buffer state deviates from the target buffer level. It turns out that this function stops buffer 1 from sending further traffic to buffer 2 when the latter reaches a certain level. Instead the excessive traffic is queued at buffer 1, since it generates lower costs.

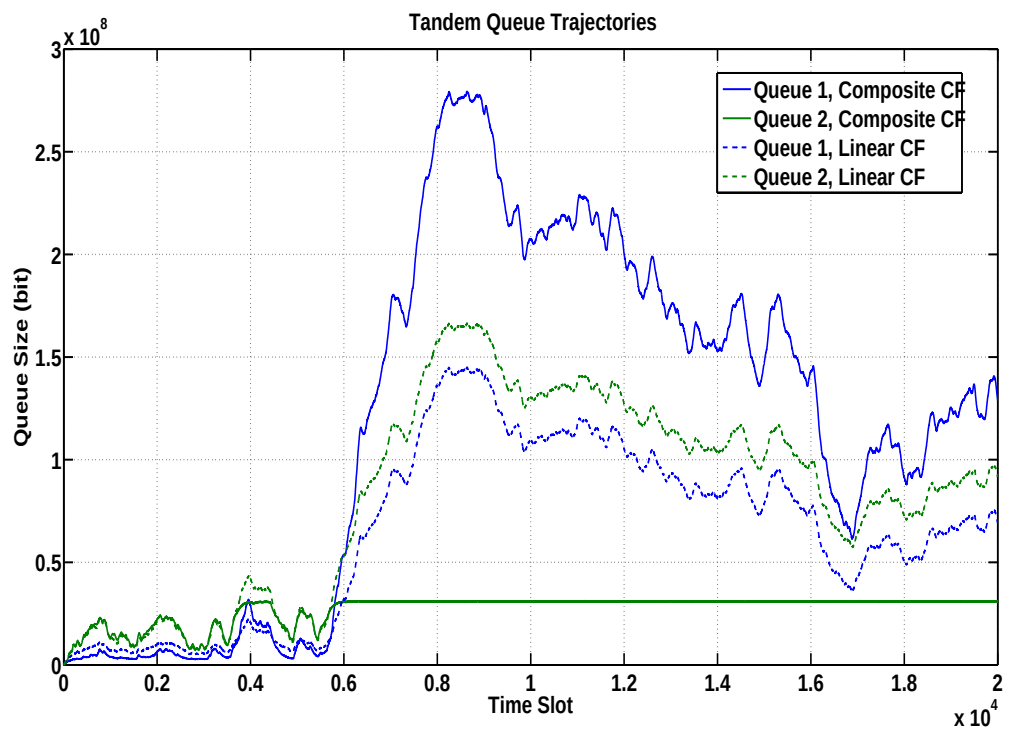

Fig. 3. Queue Trajectories under different cost-functions

[3]To increase readability we performed an exponential averaging of the (strongly fluctuating) buffer values over a time window of 100 time slots.

## VI. Design and control of large multimedia networks: numerical performance results

Using numerical experiments we subsequently evaluate the previously introduced cost functions in a larger network designed for entertainment purposes, thus especially relying on streaming traffic. Figure 4 schematically depicts the considered network.

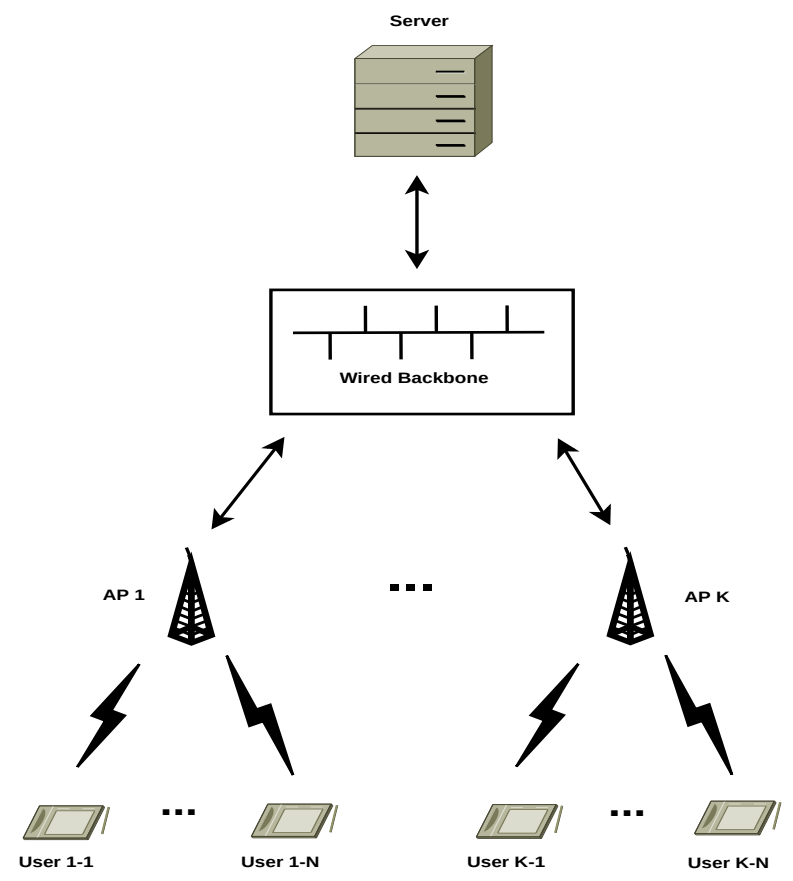

Fig. 4. Schematic representation of considered multimedia network

We have several wireless access points each serving a certain number of user terminals, potentially running streaming-based applications, in its area. Some terminals which are located in between several access points can potentially be served by more than one access point. The access points are connected via a wired backbone network to a central application server. The server itself may in turn be connected to the Internet, thus traffic for the each user arrives in a random fashion. Each component in the system has a number of queues or buffers with different requirements. The considered system can for example be used to model a wireless entertainment system (for example inside an aircraft cabin [20]).

To account for the anticipated multimedia applications we define a notion of queue outage as a measure of a policy's performance (in addition to the average cost incurred by the application of a particular control policy). Therefore we first define a buffer underflow as the event when the size of a buffer falls below a predefined value $Q^{(1)}$. The buffer underflow frequency $F_i^{\text{min}}(T)$ of user $i$ is consequently defined as the sum of the timeslots $t \leq T$ in which its buffer is lower than $Q^{(1)}$. Similarly we can define a buffer overflow event as an event where the buffer grows beyond a certain value $Q^{(2)}$. The buffer overflow frequency $F_i^{\text{max}}(T)$ of user $i$ measures the number of timeslots in which its buffer is larger than $Q^{(2)}$. The total sum of buffer outages is consequently $\bar{F}^{\text{out}}(T) = \sum_{i \in \mathcal{I}_u} F_i^{\text{min}}(T) + F_i^{\text{max}}(T)$, and eventually the relative frequency of queue outage events is defined as

$$\bar{P}^{\text{out}}(T) = \frac{1}{T \cdot I} \bar{F}^{\text{out}}(T), \qquad (18)$$

where $I$ denotes the number of user buffers. The values of $Q^{(1)}$ and $Q^{(2)}$ can be flexibly adopted to the requirements of the desired application. The goal of the control policy is consequently to keep the buffer states in between $Q^{(1)}$ and $Q^{(2)}$. Thus it is reasonable to choose $\tilde{Q} = \frac{1}{2}\left(Q^{(1)} + Q^{(2)}\right)$ for the target buffer state.

Using this definition we want to evaluate the cost functions from Section V with respect to two main aspects. First we want to ask how the queue outage performance evolves with respect to varying traffic intensities. This gives us an estimate of the robustness of the various cost functions. Second we demonstrate how the proposed approach can be used as a tool for wireless network design. We clarify this by an example, evaluating how the performance evolves with varying system bandwidths. This allows an a priori assessment of how much bandwidth has to be provided in order to support service constrains expressed by a predefined queue outage probability.

Traditionally the CRW network model summarized in Section II comprises static links. However in our system we assume time-varying wireless links between access points and terminals. Since we are mainly concerned with MAC-layer performance we use a simple abstraction of the wireless link capacities. For this, we apply a result from [21] which determines the mutual information distribution of a multi-antenna OFDM-based wireless system. To obtain rate expressions we use the notion of outage capacity (also derived in [21]) $\mathcal{I}_{\text{out},p_o}$ for a given outage probability $p_o$, defined as the maximum rate guaranteed to be supported for $100(1-p_o)\%$ of the channel realizations and is given by

$$\mathcal{I}_{\text{out},q} = \mathbb{E}[\mathcal{I}_{\text{OFDM}}] - \sqrt{Var(\mathcal{I}_{\text{OFDM}})}Q^{-1}(p_o)$$

with $Q(\cdot)$ being the Gaussian Q-function and $p_o$ the desired outage probability. $\mathbb{E}[\mathcal{I}_{\text{OFDM}}]$ and $Var(\mathcal{I}_{\text{OFDM}})$ are determined according to above mentioned mutual information distribution.

In the following we show performance results in an example network structured according to Figure 4. Configuration details of the simulations are summarized in Table I.

TABLE I
General Simulation Parameters

| Parameter | Value |
|---|---|
| Simulation Duration ($T$) | 100000 time slots |
| Number of Users per AP | 10 |
| Number of Access Points | 3 |
| Wired Link Capacity ($R_s$) | 100 Mbit/s |
| Wireless Link Outage Probability ($p_o$) | 0.01 |
| Target User Buffer Size ($\tilde{Q}$) | 20 Mbit |
| Minimum User-Buffer Constraint ($Q^{(1)}$) | 10 Mbit |
| Maximum User-Buffer Constraint ($Q^{(2)}$) | 30 Mbit |
| Application Rate ($R_a$) | 3 Mbit/s |
| Iterations of Pick-And-Compare ($n_s$) | 100 |

## A. Traffic Intensity

Especially when traffic is generated outside of the considered network it is reasonable to assume that the offered arrival rate can deviate from the anticipated operating point. Therefore we subsequently investigate how the candidate cost functions influence the policy's behaviour with varying traffic intensity. Note, that the application drains application queues at a rate of 3Mbit/s, thus intuitively one can expect that at traffic lower than this value the influence of underflows dominates while at traffic rates larger then this value overflows are more likely to occur. We first consider the buffer underflow probability, since as opposed to other state-of-the art control approaches, a main incentive considered here is to prevent service interruptions due to low buffers. Consider Figure 5, comparing the underflow probability of various cost functions together with classical MaxWeight as baseline. When the mean arrival rate equals the application service rate $R_a$ (or is larger) all policies produce low underflow frequencies, however one can already observe a performance gain from the more sophisticated cost functions (CF). The gain significantly grows when the arrival rates are slightly lower than $R_a$. While MaxWeight and the linear CF (13) show almost the same high underflow frequency (since both controls do not penalize low buffer states at the application buffers), already the shifted quadratic CF (14) shows some improvement. The best performance is obtained with the sophisticated bandstop and composite CFs (given by (16) and (15), respectively) which assign different costs to application buffers and non-application buffers. However, the performance gap between the two is relatively small. Therefore we choose the much simpler CF (15) in the following. However not only buffer underflows have to be avoided, in some cases it may be desirable to avoid large queues as well. Therefore we next investigate the performance with respect to the queue outage probability defined in (18). Figure 6 summarizes the results. Naturally, at traffic arrival rates lower than the application's service rate, all cost functions produce decreasing outages with increasing arrival rates, since queue underflows become less frequent. Beyond the applications service rate, the policies without a sophisticated cost function rapidly increase the queue outages with further increasing traffic due to overflows. Only the composite cost function can further decrease the number of queue outages, since exceeding traffic is stored at buffers that produce lower costs (cf. the tandem network in Section V-A).

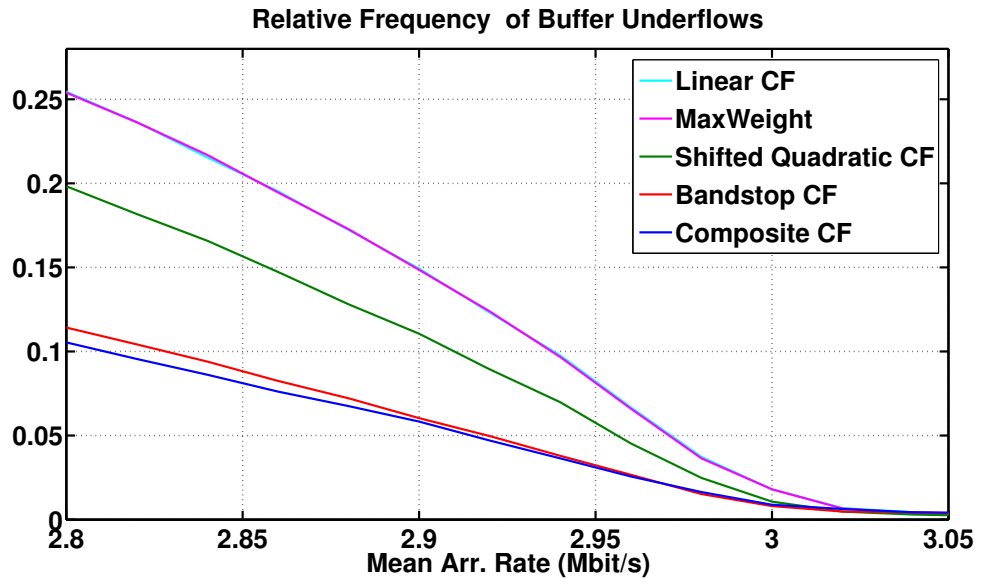

Fig. 5. Buffer underflow frequencies obtained by different cost functions

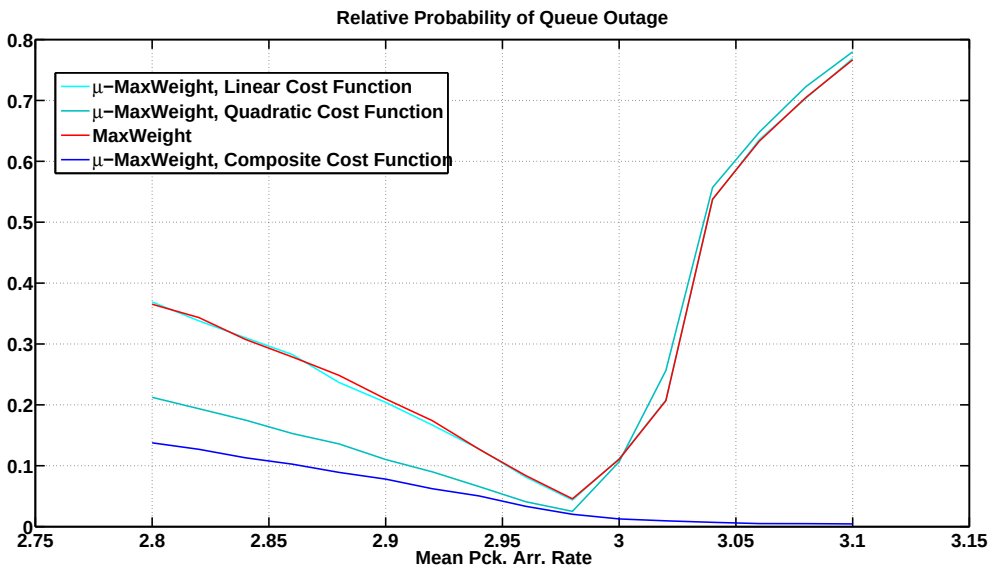

Fig. 6. Relative queue outage frequencies of various cost functions

### *B. Network Design*

Given a concrete application which requires a certain outage performance, it is of great interest how many system resources have to be provided. For example when planning some entertainment network it is important to know how much wireless bandwidth is actually needed to support the desired services. Therefore we next investigate the influence of the system bandwidth, when a certain target outage rate is to be achieved. For this, we use the composite cost function (15), and again plot MaxWeight as a baseline. Figure 7 compares the relative frequency of user buffer outages for varying wireless bandwidths. Assume we want a queue outage probability (defined according to (18)) of 2% not to be exceeded, marked by the dashed black line in Figure 7. Obviously a simple scheduling strategy which is not designed based on a cost-function such as MaxWeight is not able to push the outage performance below this limit. By contrast using the cost-function based control a wireless bandwidth of 20 MHz would be sufficient to guarantee the desired performance in the considered configuration.

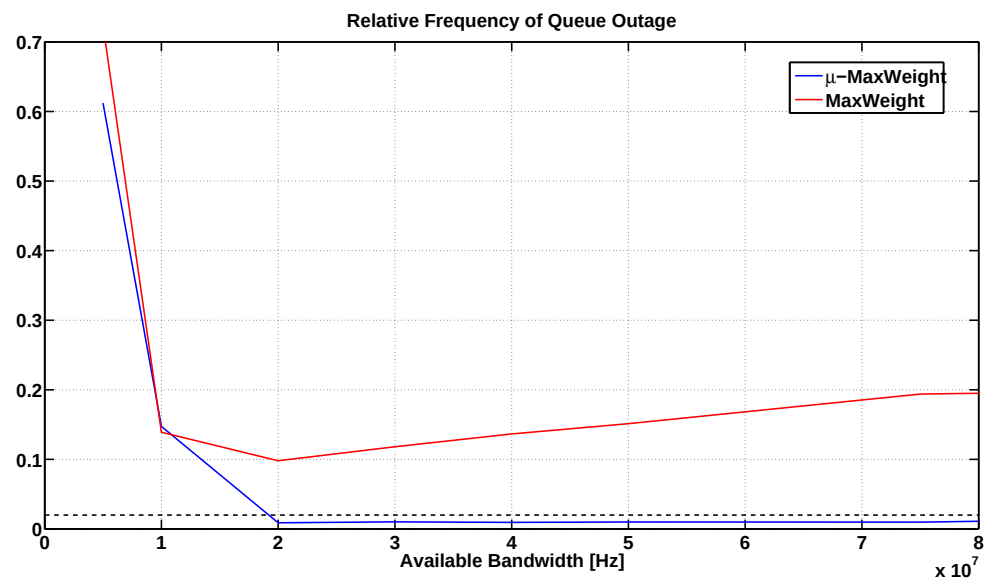

Fig. 7. Comparison of policies with respect to system bandwidth

## VII. Conclusions

We presented a framework for the design and control of queueing networks with special emphasis on service induced constraints. Therefore we introduce a cost-function based approach to queueing control and provide sufficient conditions for its throughput optimality. The inherent complexity is tackled by a randomization approach. Furthermore we show how particular buffers can be steered to desired operating points with an appropriate cost function design, which can be used for example to avoid buffer underflows in a streaming-traffic based scenario. We evaluate different cost functions for their suitability using numerical simulations of a large entertainment network. Eventually we demonstrated how the approach can aid the network design process, by a priori determining how many system resources are needed to support given service requirements.